# Pulse-duration dependence of saturable and reverse saturable absorption in $ZnCo_2O_4$ microflowers

Pritam Khan[1, 2], Rajesh Kumar Yadav[1], Anirban Mondal[1], Chandra Sekhar Rout[3] and K. V. Adarsh[1*]

[1]Department of Physics, Indian Institute of Science Education and Research,

Bhopal 462066 India

[2]Department of Physics and Bernal Institute, University of Limerick, Limerick V94T9PX,

Ireland

[3]Centre for Nano and Material Sciences, Jain University, Ramanagaram, Bangalore, 562112,

India

*We employed open-aperture Z-scan technique to unveil the third-order optical nonlinearity in $ZnCo_2O_4$ (ZCO) microflowers. Our results indicate that intersystem crossing (ISC) lifetime can be used as simple tool to demonstrate remarkably contrasting optical nonlinearity in ZCO. Ultrafast transient absorption measurements reveal that ISC from singlet to triplet state takes place in 5 ps. For femtosecond laser pulses, when the pulse duration is shorter than ISC lifetime, saturable absorption (SA) takes place for all intensities. On the contrary, when the pulse duration is longer than ISC for nanosecond excitation, we observe transition from SA to reverse SA (RSA) at higher intensities via excited-state absorption. We envisage that benefiting from SA and RSA, ZCO emerges as potential candidate for mode locking and optical limiting devices.*

* Author to whom correspondence should be addressed. Electronic mail: adarsh@iiserb.ac.in.

# 1. Introduction

It is well known that nonlinear optical responses in the form of saturable absorption (SA) and reverse saturable absorption (RSA) offers tremendous potential application in mode locking and optical limiting, respectively [1, 2]. Likewise, there has been much interest in developing new nonlinear optical materials for the fabrication of next generation opto-electronic devices. Till date, several materials, including chalcogenide glasses (ChGs) [3, 4], graphene oxide [5], transition metal dichalcogenides ($MoSe_2$, $WS_2$) [6, 7] II–VI compound semiconductors (ZnO, $LiNbO_3$, $LiTaO_3$) have been studied extensively for numerous potential applications [8-10]. Among these materials, ZnO stands apart thanks to its high nonlinear refractive index ($n_2$) and absorption coefficient ($\beta$) and strong second ($\chi(2)$) and third order ($\chi(3)$) optical susceptibilities, that is ideal for optoelectronics [8, 11, 12]. Apart from that ZnO has added advantages of high excitonic gain, large exciton binding energy (60 meV), direct wide band gap (3.3 eV), with high mechanical and chemical stability, non-toxicity, and natural abundance [11, 12].

In quest for enhancing the nonlinearity further and broaden the application boundary it is quite tempting to hybridize ZnO with transition metals like Co to facilitate Zinc Cobaltite Oxide ($ZnCo_2O_4$, hereafter ZCO). It is already well established that ZCO offers tremendous potential application in diverse field of Li-ion batteries [13], gas-sensors [14], supercapacitors [15] etc. other than optics. However, ZCO exhibits high optical transparency over broadband wavelength regime and process electrical conductivity and therefore emerged as well-known transparent conducting oxides (TCOs) material [16]. ZCO microstructures with flower-like features are of particular important because thin edges and sharp tips associated with such geometry could provide favourable enhancement of the optical field [17, 18]. Despite having high linear optical transmission in UV-Vis-NIR regime with the advantage of wide bandgap, the nonlinear optical response of ZCO microstructures is hitherto unexplored.

In this regard, here we studied for the first time the nonlinear optical response of ZCO microflowers by performing Z-scan measurements with femtosecond (fs) and nanosecond (ns) laser. ZCO exhibits remarkably contrasting nonlinear absorption depending on the intersystem crossing lifetime and pulse duration of the excitation laser. Precisely, for fs laser excitation, only SA takes place, whereas when excited with ns laser, ZCO exhibits a transition from SA to RSA at higher intensities. By considering five-level energy diagram, we showed that SA takes place between ground and first singlet state, whereas RSA originates from excited-state absorption (ESA) between the triplet states.

## 2. Methods

*2.1 Sample preparation and structural characterization*

We synthesized ZCO microflowers by using low cost, low temperature hydrothermal route and the details can be found elsewhere [17, 18]. Precisely, under hydrothermal conditions, supersaturation takes place between the co-precipitated particles, e.g., Zn and Co, which favors the formation of nanosheet-like structures. Finally, these nanosheets self-assembled to form thermally stable tiny spherical structures. Morphology and size of the ZCO microflowers are characterized by high resolution field emission scanning electron microscope (HR FESEM, Zeiss, ULTRA Plus). The crystallinity of the sample is characterized by X-ray diffraction (XRD, PANalytical Empyrean) Cu-Kα radiation ($\lambda$ = 1.54184 Å) with a step size of 0.02°.

*2.2 Raman spectroscopy*

We recorded the Raman spectra in a Horiba JY LabRam HR Evolution Spectrometer mounted with a grating of 1800 grooves/mm using a 50 X objective with N.A. = 0.5 lens in back-scattering geometry. We excite the microflowers by a 532 nm diode laser. The detection is performed by an air-cooled charge coupled device (CCD) detector. The measurements are performed at very nominal power of ~ 1 mW to avoid any lightinduced effects.

*2.3 Ultrafast pump-probe measurements*

For femtosecond pump-probe measurements, the fundamental beam of 120 fs pulses cantered at 800 nm with a repetition rate of 1 kHz was split into two beams to generate pump and probe [25]. We used 400 nm pump beam of fluence 0.25 mJ/cm$^2$. On the other hand, white light continuum (450-850 nm) of very low fluence (6 μJ/cm$^2$) was used as the probe beam. We define IA of the probe beam $\Delta A = \log [I_{ex}(p)/I_0(p)] - \log [I_{ex}(r)/I_0(r)]$. The symbols *p* and *r* correspond to the probe and the reference, respectively. $I_{ex}$ and $I_0$ represent transmitted probe intensities at delay time *τ* following the excitation of the pump beam and in the ground state, respectively.

*2.4 Z-scan measurements*

To explore the nonlinear optical response of ZCO microflowers, we employed conventional open aperture Z-scan technique. We performed two sets of Z-scan measurements: First with 532 nm, 7 ns pulses from the second harmonic of a Nd-YAG laser and second using 560 nm, 120 fs pulses from the Ti-sapphire laser that excite the sample. We kept the repetition rate of the laser at 10 Hz to prevent the sample from heating and photo damage. The laser beam was focused by a plano-convex lens of focal length 20 cm while the sample was moved by computer-controlled translation stage.

## 3. Results and Discussions

The SEM images of the synthesized ZCO microflowers are shown in Fig. 1(a). The SEM images reveal uniform flower-like microspherical structures with diameter between 6-10 µm. Fig. 1(b) shows the XRD pattern of the ZCO, indicating preferential growth along (3 1 1) direction. The diffraction peaks are indexed to cubical spinel structured ZCO (JCPDS NO: 23-1390). The Raman spectrum of ZCO shown in Fig. 1(c) consists of four dominant bands. The peak at 472 cm$^{-1}$ is associated with strong vibration due to the stretching of the bonds between Co–O and Zn–O and assigned to $E_g$ mode. The moderately intense peak at 518 cm$^{-1}$ belongs to $F_{2g}^{(2)}$ symmetry, from the stretching of the Co–O bond. The weak shoulder at 611 cm$^{-1}$ corresponds to $F_{2g}^{(1)}$ symmetry. The strongest of the Raman band at 673 cm$^{-1}$ is arising from the relative contribution of two peaks at 658 and 682 cm$^{-1}$, both having A$_{1g}$ symmetry [17, 18]. To determine the bandgap, we showed in Fig. 1(d) the optical absorption spectra of ZCO. Clearly, the spinel ZCO microflowers exhibits strong absorbance from 300 to 1000 nm, spreading over UV to IR wavelength regime. ZCO being the direct bandgap material, we calculate the optical bandgap using the Tauc equation [19] expressed as:

$$(\alpha h\nu)^2 = (h\nu - E_g) \qquad (1)$$

Where, $h\nu$, $\alpha$ and $E_g$ are the photon energy, absorption coefficient and bandgap respectively. From the best fit to the experimental data (Inset), the optical bandgap is found to be 1.72 and 2.73 eV, in accordance with the work by Guo [20] and Yu [21].

To obtain the intersystem crossing lifetime for ZCO, we perform femtosecond pump-probe TA measurements [22]. The TA spectra is measured from 1.65 to 2.25 eV for 400 nm excitation wavelength up to a probe delay of 600 ps. SA and RSA in Z-scan measurements is generally evinced by the absorption bleaching (negative TA) and induced absorption (positive TA), respectively. It can be seen from Fig. 2(a) that the positive TA signal starts developing from 0.3 ps and reaches maximum at 1 ps. With increase in delay, TA decays and saturates within 200 ps. It is quite evident that different regions of TA spectra decays at different time. In this context, relaxation kinetics of TA signal is evaluated by looking into the temporal evolution of TA at three contrasting energies as depicted in Fig. 2(b). For all probe energies, TA signal is fitted well by a two-component exponential decay model with time constants summarized in Table 1. Clearly, lower energy component of TA decays faster in line with previous observations [23, 24].

The faster component ($\tau_1$) takes place within 10s of ps, which we assign to intersystem crossing (ISC) process from first singlet state $S_1$ to first triplet state $T_1$ as shown in the five-level energy diagram [25-27] in Fig. 3. On the other hand, the slower component ($\tau_2$) is associated with very long triplet ($T_1$) lifetime. Such inference is further substantiated from fig. 2(b) that TA is not completely reversible as it does not revert to zero value even after long delay of ca. 600 ps. The five-level model shown in Fig. 3 consists of ground state $S_0$; singlet states $S_1$, $S_2$; and triplet states $T_1$, $T_2$. Optical transition between these levels determines the nonlinear absorption that will take place in the sample. For example, direct transition between $S_0$ to $S_2$ leads to instantaneous two-photon absorption (TPA) whereas sequential transition between $S_0 \rightarrow S_1 \rightarrow S_2$ is responsible for two-step resonant TPA. On the other hand, ESA takes place between either singlet ($S_1 \rightarrow S_2$) or triplet ($T_1 \rightarrow T_2$) excited states depending on the intersystem crossing lifetime. We will show in the following section on how the transition between different energy levels give rise to various nonlinear process in ZCO. Likewise, SA and ESA can be differentiated in terms of the absorption cross sections of ground state, $S_0 \rightarrow S_1$ ($\sigma_{gs}$) and excited state $S_1 \rightarrow S_2$ or $T_1 \rightarrow T_2$ ($\sigma_{es}$). Specifically, the ratio $\sigma_{es}/\sigma_{gs} < 1$ leads to SA, whereas $> 1$ results in ESA. In this context, we calculate $\sigma_{gs}$ and $\sigma_{es}$ using following equations [28]:

$$\sigma_{gs} = \frac{-\log T_0}{NL} \qquad (2)$$

$$\sigma_{es} = \frac{-\log T_{max}}{NL} \tag{3}$$

where $T_0, T_{max}$, $N$, and $L$ are the transmission in the linear regime, the high intensity saturated transmission, ground state carrier density and thickness of the ZCO microflower.

To exploit the ISC lifetime, first we performed the conventional open-aperture Z-scan measurements [3, 4, 29] with 560 nm femtosecond laser where the pulse-width of the laser $\tau_{fs}$ ~120 fs is shorter than the ISC process ($\tau_{ISC}$ ~ 5 ps). Fig. 4(a) shows the open-aperture Z-scan traces under 560 nm, 120 fs laser excitation at peak intensities of 89 (green), 122 (blue), 209 (red) GW/cm$^2$ at the focal point. The reference signal is obtained from the pure ethanol solution (magenta), without the dispersion of ZCO microflowers. A flat line is observed before and after the focal point, indicating that ethanolic solution does not contribute to the nonlinear response of ZCO. The peak shape of the Z-scan traces indicates that ZCO exhibits SA at all three intensities when excited with fs laser. Experimental observation is further supported by the ratio $\sigma_{es}/\sigma_{gs}$ which remains 0.92, 0.88 and 0.82 (all <1) at 89, 122, 209 GW/cm$^2$ in line with SA. The SA here can be physically explained from the energy-level diagram of Fig. 3. Upon excitation, the carriers first rise from $S_0$ to $S_1$, but because of the shorter pulsed duration of $\tau_{fs}$ ~120 fs, instead of going to $T_1$ via ISC, the carriers decay to $S_o$ directly. Therefore, SA for fs illumination can be explained by an effective two-level model instead of five-level one. Likewise, in the TA spectra shown in Fig. 2(a), we could not observe absorption bleaching associated with SA observed in fs Z-scan measurements, because such process takes place in extremely faster time scales within 100s of fs that is limited by the instrument response function (IRF) of the pump-probe TA set up.

The Z-scan theory is employed to quantify SA and ESA by following propagation equation in the dispersion as a function of the position given by [30]:

$$\frac{dI}{dx} = -\alpha(I)I \tag{4}$$

Where $I$, $x$, and $\alpha(I)$ are the intensity, propagation distance in the sample and intensity dependent absorption coefficient respectively. $\alpha(I)$ is generally defined by:

$$\alpha(I) = \frac{\alpha_0}{1 + I/I_s} + \beta_{ESA} I \qquad (5)$$

Where $\alpha_0$, $I_s$ and $\beta_{ESA}$ are the linear absorption coefficient, saturable intensity, and ESA coefficient, respectively. By using Eqns. [4] and [5] we calculate $I_s$ to be 600, 590, and 643 GW/cm$^2$, respectively at 89, 122, 209 GW/cm$^2$. The $I_s$ of ZCO is also found to be comparable with the conventional saturable absorber e.g., MoSe$_2$ (590 GW/cm$^2$) and graphene (583 GW/cm$^2$) [31]. The constancy in $I_s$ at different intensities confirms that ZCO is a good saturable absorber and a potential candidate for passive mode locking in picosecond and femtosecond pulse generation. To further demonstrate practical applications of our system, we have plotted in Fig. 4(b) the variation of output intensity ($I_{out}$) as a function of input intensity ($I_{in}$) at peak intensities of 209 GW/cm$^2$. Clearly, $I_{out}$ scales linearly with $I_{in}$ for lower values, following Lambert-Beer law. However, as $I_{in}$ increases, $I_{out}/I_{in}$ deviates from linearity as depicted by the dashed line. This particular value of $I_{in}$ is known as the threshold intensity which plays an important role for mode locking.

Next, we performed second set of Z-scan measurements with 532 nm nanosecond laser and in this case the laser pulse width $\tau_{ns}$ ~ 5 ns is much longer than $\tau_{ISC}$ ~ 5 ps. Fig. 5(a) shows the open-aperture Z-scan traces of ZCO microflowers when excited with 5 ns laser with peak intensities: 0.10 (green), 0.21 (blue) and 0.37 (red) GW/cm$^2$, measured at the focal point. At lowest intensity, the Z-scan curve shape indicates that the sample undergoes SA. With increase in intensity to 0.21 GW/cm$^2$, we observed weak RSA within the SA curve, whereas at 0.37 GW/cm$^2$, the sample exhibits strong RSA. Thus, we observed an intensity dependent crossover from SA to RSA in ZCO. In general, RSA can take place either by two-photon absorption (TPA) or excited state absorption (ESA). In the present case, as the excitation energy is higher than the bandgap energy of ZCO, we can exclude the possibility of TPA, which leaves us with ESA. SA and ESA again here can be explained from the five-level model shown in Fig. 3. At low intensity, carries from $S_0$ are first excited to first singlet state $S_1$ and decays immediately to first triplet state $T_1$ by intersystem crossing within few ps. At 0.10 GW/cm$^2$, the intensity is not strong enough to excite the carriers further from $T_1$ to $T_2$, and carriers remain in $T_1$ for a longer time ($\tau_2$) before they eventually decay to $S_0$. Using eqns. [2] and [3] the ratio $\sigma_{es}/\sigma_{gs}$ is found to be 0.63 (<1) thus confirms the observation of SA at low intensity. The RSA at 0.37 GW/cm$^2$ takes place when the carriers following the decay from $S_1$ to $T_1$ by ISC, further excited to second triplet state $T_2$ owing to higher intensity. Consequently, at 0.37 GW/cm$^2$ the ratio $\sigma_{es}/\sigma_{gs}$ gives us 1.33 (>1), characteristic of ESA. From Eqns. [4] and [5] we found that $I_s$ are 0.04 ± 0.003, 0.03 ± 0.004, 0.04 ± 0.004 GW/cm$^2$ for peak intensities 0.10, 0.21, 0.37 GW/cm$^2$, respectively. It is apparent that the $I_s$ values remain nearly constant at these intensities thus indicates the quality of ZCO as a good saturable absorber. On the other hand, $\beta_{ESA}$ is found to be 27.4±3.1 and 129±19 cm/GW at 0.21, 0.37 GW/cm$^2$, respectively. Clearly, ESA coefficient increases remarkably at higher intensities in line with Fig. 5(a). Observation of SA at lower intensity and crossover to ESA at higher intensity highly advantageous for the fabrication of nanophotonic devices. For example, SA is exploited for passive mode locking [3] whereas ESA finds application in optical limiting [6]. In Fig. 5(b), we have shown the output intensity as a function of the input intensity for ZCO at the at the peak intensity of 0.37 GW/cm$^2$ which shows a linear behaviour at low input intensity and deviation from linearity, i.e., the output intensity is limited by the material for all higher input intensities, i.e., our system becomes active optical limiter.

In summary, we successfully exploit the interplay between intersystem crossing and laser pulse duration to demonstrate contrasting nonlinear absorption in ZCO microflowers. From Z-scan measurements, we found that for fs excitation, when the pulse duration of laser is shorter than ISC lifetime, SA takes place between ground and first singlet state. On the other hand, when the pulse duration is longer than ISC process for ns excitation, we observed a transition from SA to RSA at higher intensity thanks to ESA from triplet states. We found that ESA coefficient ($\beta_{ESA}$) of ZCO is ~ 15 times stronger than that of conventional ZnO, whereas saturation intensity ($I_s$) associated with SA is comparable with the well-known saturable absorber like graphene and $MoSe_2$. We foresee that the observation of SA and ESA together in ZCO opens up new avenue for its application in mode locking and optical limiting devices.

## Acknowledgements:

The authors gratefully acknowledge the Department of Science and Technology (Project no: EMR/2016/002520), DAE BRNS (Sanction no: 37(3)/14/26/2016-BRNS/37245). CSR acknowledges financial support from Department of Science and Technology (DST)-SERB Early Career Research project (Grant No. ECR/2017/001850), DST Nanomission (DST/NM/NT/2019/205(G)) and Karnataka Science and Technology Promotion Society (KSTePS/VGST-RGS-F/2018-19/GRD NO. 829/315).

## Data availability:

The data that support the findings of this study are available from the corresponding author upon reasonable request.

**Table 1.** Decay time constants associated with the transient absorption

| Energy (eV) | $\tau_1$ (μs) | $\tau_2$ (μs) |
| --- | --- | --- |
| 1.75 | 4.2 ± 0.5 | 234 ± 29 |
| 1.95 | 5.1 ± 0.6 | 286 ± 31 |
| 2.20 | 16.3 ± 2.1 | 329 ± 42 |

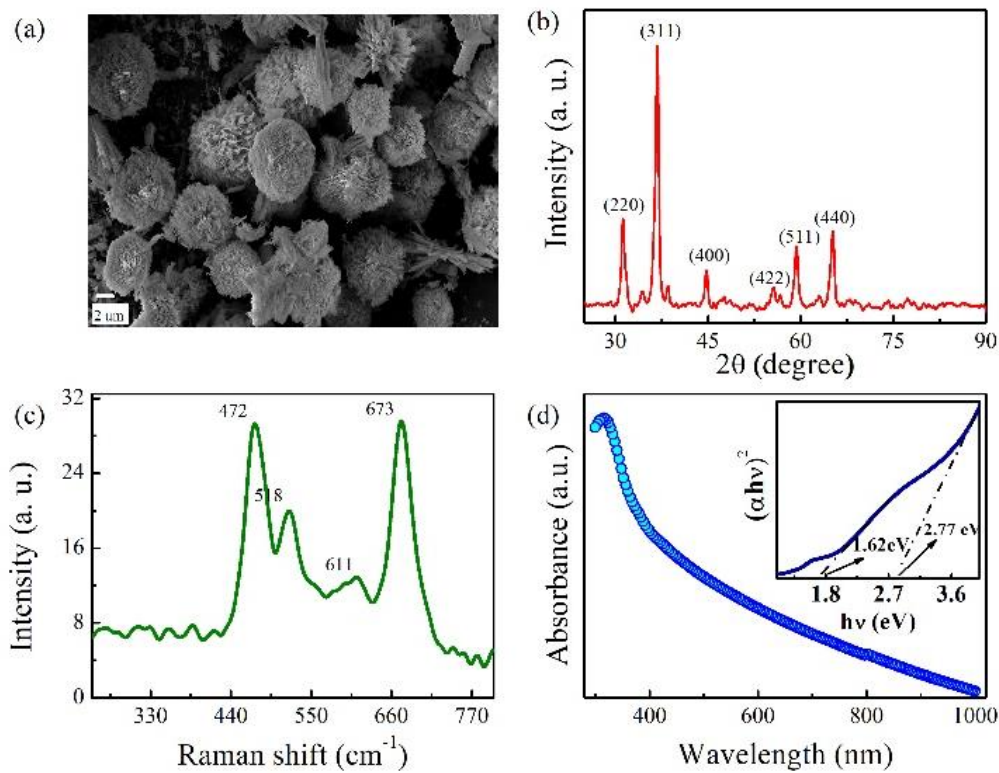

**Fig. 1.** (a) SEM image of ZCO revealing uniform flower-like micro-spherical structures with diameter between 6-10 µm. (b) XRD and (c) Raman spectrum ZCO of microflowers. (d) Optical absorption spectrum of the ZCO microflowers dispersed in ethanol. The inset represents the Tauc plot for calculating optical bandgap.

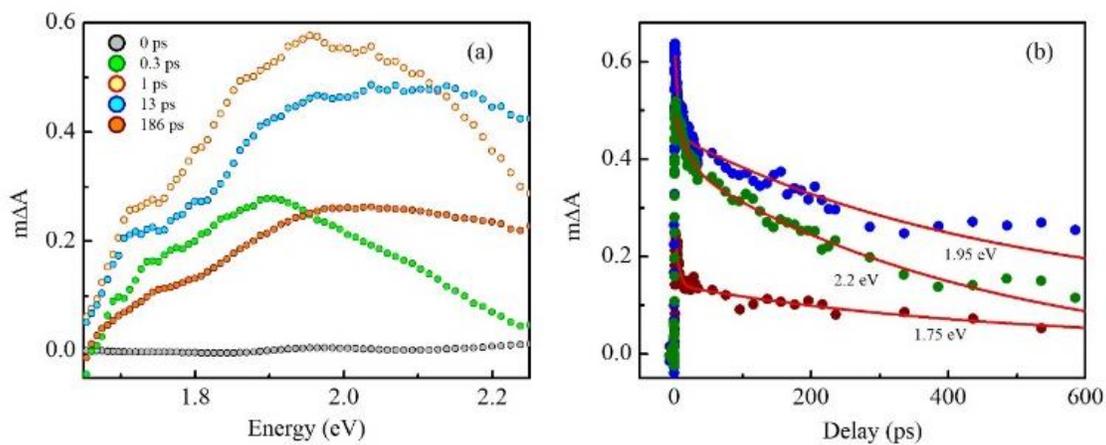

**Fig. 2.** (a) Ultrafast TA spectra of ZCO microflowers at different probe delays. (b) Temporal evolution of TA at three different energies.

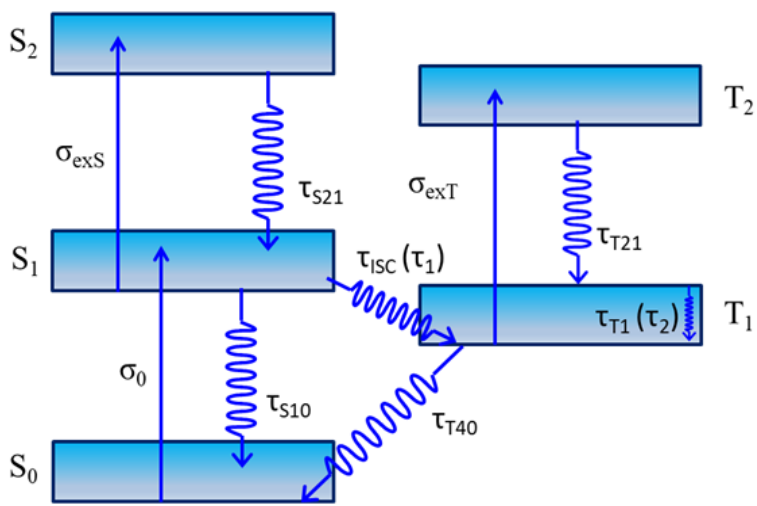

**Fig. 3.** Schematic of the five-level energy diagram with ground state $S_0$; singlet states $S_1$, $S_2$; and triplet states $T_1$, $T_2$. Solid upward arrows imply excitation resulting from photon absorption and dashed curved downward arrows represent relaxations.

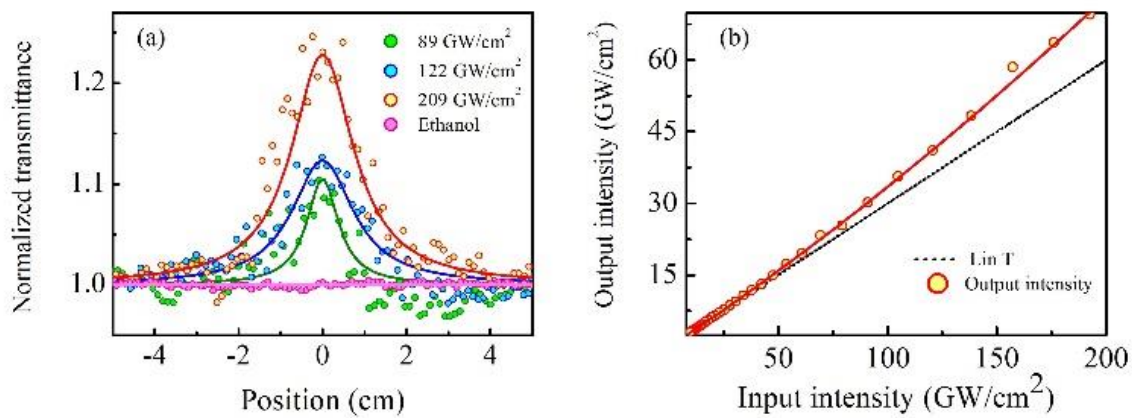

**Fig. 4.** (a) Open aperture Z-scan traces at various intensities when excited with 120 fs, 560 nm laser pulses. (b) Variation of output intensity as a function of input intensity for ZCO at the peak intensity of 209 GW/cm$^2$.

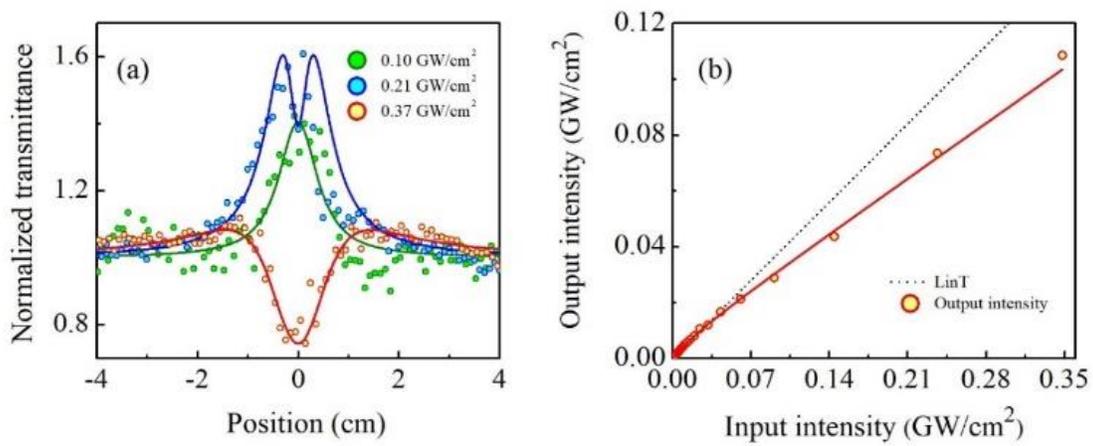

**Fig. 5.** Open aperture Z-scan traces of ZCO at different pumping intensities when excited with 5 ns, 532 nm laser pulses. (b) Variation of output intensity as a function of input intensity for ZCO at the peak intensity of 0.37 GW/cm$^2$.